\documentclass[amssymb,useAMS,prd,aps,amsmath,superscriptaddress,nofootinbib,twocolumn]{revtex4}
\usepackage{color}
\usepackage{pslatex}
\usepackage{pdfpages}
\usepackage{graphicx}
\usepackage{psfrag}
\usepackage{pdftricks}
\usepackage{multirow}

\begin{document}
\newcommand{\addCorr}[1]{{#1}}
\newcommand{\deu}{D}
\newcommand{\tro}{$^3$He}
\newcommand{\qua}{$^4$He}
\newcommand{\six}{$^{6}$Li}
\newcommand{\sep}{$^{7}$Li}
\newcommand{\be}{$^{7}$Be}
\newcommand{\neu}{$^{9}$Be}
\newcommand{\dix}{$^{10}$B}
\newcommand{\onz}{$^{11}$B}
\newcommand{\hli}{$^4$He, D, $^3$He and $^{7}$Li}
\newcommand{\sbbn}{Standard Big-Bang Nucleosynthesis}
\def\la{\lower.5ex\hbox{$\; \buildrel < \over \sim \;$}}
\def\ga{\lower.5ex\hbox{$\; \buildrel > \over \sim \;$}}
\newcommand{\mnras}{Mon.Not.Roy.Astron.Soc.}
\newcommand{\aap}{Astron.Astrophys.}

\title{Neutron injection during primordial nucleosynthesis alleviates the primordial \sep\ problem}

\author{Daniel Albornoz V\'asquez}
\affiliation{Institut d'Astrophysique de Paris, UMR 7095 CNRS, Universit\'e Pierre et Marie Curie, 98 bis Boulevard Arago, Paris 75014, France}

\author{Alexander Belikov}
\affiliation{Institut d'Astrophysique de Paris, UMR 7095 CNRS, Universit\'e Pierre et Marie Curie, 98 bis Boulevard Arago, Paris 75014, France}

\author{Alain Coc}
\affiliation{Centre de Spectrom\'etrie Nucl\'eaire et de Spectrom\'etrie de Masse (CSNSM), CNRS/IN2P3, Universit\'e Paris Sud 11, UMR~8609, B\^atiment 104, F--91405 Orsay Campus, France}

\author{Joseph Silk}
\affiliation{Institut d'Astrophysique de Paris, UMR 7095 CNRS, Universit\'e Pierre et Marie Curie, 98 bis Boulevard Arago, Paris 75014, France}
\affiliation{Beecroft Institute of Particle Astrophysics and Cosmology, 1 Keble Road, University of Oxford, Oxford OX1 3RH UK}
\affiliation{Department of Physics and Astronomy, 3701 San Martin Drive, The Johns Hopkins University, Baltimore MD 21218, USA}

\author{Elisabeth Vangioni}
\affiliation{Institut d'Astrophysique de Paris, UMR 7095 CNRS, Universit\'e Pierre et Marie Curie, 98 bis Boulevard Arago, Paris 75014, France}

\date{\today}

\begin{abstract}
We present a parametrized study of the effects of free thermal neutron injection on primordial nucleosynthesis, where both the rate and the time scale of injection are varied. This generic approach is found to yield a successful solution for reducing the \sep\ abundance without causing significant problems to other elemental abundances. Our analysis demonstrates that hadronic injection, possibly due to decays or annihilations of dark matter particles with a mass of about 1 to 30 GeV, provides a possible solution to an outstanding problem in the standard Big Bang model.
\end{abstract}

\maketitle

\section{Introduction}
The motivation for this study concerns the discrepancy between the primordial Li abundance predicted in the canonical Big Bang model and observational data. The primordial lithium abundance is deduced from observations of low metallicity stars in the halo of our Galaxy where the lithium abundance is almost independent of metallicity, displaying a plateau, the so-called Spite plateau \citep{Spite82}. This interpretation assumes that lithium has not been depleted at the surface of these stars, so that the presently observed abundance is supposed to be equal to the initial value. The small scatter of values around the Spite plateau is an indication that depletion may not have been very effective. Astronomical observations of these metal-poor halo stars \citep{Ryanetal2000} have led to a relative primordial abundance of:

Li/H~$ = (1.23^{+0.34}_{-0.16}) \times 10^{-10}$.

\noindent A more recent analysis by \citet{sbordone10} gives:
  
Li/H~$ = (1.58 \pm 0.31) \times 10^{-10}$.

\noindent More generally, \citet{Spite10} have reviewed recent Li observations and their different astrophysical aspects. Also see \citet{frebel11} for a comprehensive review.
 
On the other hand, the most recent \sbbn\ (SBBN) calculations, using the most up-to-date nuclear data, give: 

Li/H = $(5.14\pm0.50)\times 10^{-10}$ \citep{CV10}.
 
\noindent Hence there is a factor of 3-4 discrepancy between observation and theory at the WMAP7 baryonic density.

\sep\ is produced as a by-product of decay of $^7$Be. Nuclear mechanisms to destroy this $^7$Be have been explored. \addCorr{A possibly} increased $^7$Be(d,p)$2\alpha$ cross section has been proposed by \citet{Coc04} and later by \citet{Cyb09} but was not confirmed by experiments \citep{Ang05,OMa11,Kir11}. Other $^7$Be destruction channels have recently been proposed by \citet{Cha11}
and await experimental investigation. 

Another scenario would be to take advantage of an increased late-time neutron abundance, as introduced in~\cite{Reno:1987qw} for the generic case of hadronic injection. In the context of varying constants, when the $^1$H(n,$\gamma)^2$H rate is decreased, the neutron late-time abundance is increased (with no effect on \qua) so that more $^7$Be is destroyed by $^7$Be(n,p)$^7$Li(p,$\alpha)\alpha$, \citep[see in] [Fig. 1]{Coc07}. Many other nuclear reactions could be potential sources of free neutrons. However, a recent study \cite{Coc11} extended the SBBN network to 59 nuclides from neutrons to $^{23}$Na, linked by 391 reactions involving n, p, d, t and \tro\ induced reactions and 33 $\beta$-decay processes. The \sep\ abundance is now estimated to Li/H = $5.24\times 10^{-10}$ \citep{Coc11}, as found also by \citep{Cyb08}. This confirms and even increases the discrepancy.

Including physics beyond the standard model of particle physics and beyond the standard Big-Bang picture can also give rise to extra neutron injection. Indeed, BBN can be used as an anchor to test the plausibility of new physics, and conversely, new physics can provide mechanisms to help solving the SBBN discrepancies with observations~\cite{Malaney:1993ah,Sarkar:1995dd,Jedamzik:2009uy,Pospelov:2010hj}.
One such option is that of hadronic decays of exotic unstable particles. For example, a metastable stop Next-to-Lightest Supersymmetric Particle (NLSP) decays into a gravitino Lightest Supersymmetric Particle (LSP), thus a dark matter candidate, and a top quark injects energetic protons and neutrons during nucleosynthesis~\cite{PhysRevD.79.043514,1988ApJ...330..545D,Reno:1987qw,Jedamzik:2004er,Jedamzik06,Cumberbatch:2007me,Cyburt:2010vz,Pospelov:2010cw}. Another possible source of neutrons arises from residual annihilations of dark matter particles -- such as neutralino LSP annihilating into fermion-antifermion couples that further hadronize -- that are chemically decoupled at BBN times~\cite{Reno:1987qw,Jedamzik:2004er,Pospelov:2010hj}. In all these scenarios, neutron injection provides the primary impact on BBN and Li production.

\addCorr
{
In this work, we study the effect of free neutron injection, parametrized by the injection rate and time-scale. Different injection models are thus included in the full code presented in~\cite{Coc11}. Hence, this implementation is expected to give hints regarding the injection mechanism including possible nuclear reaction uncertainties, fundamental constant variations and exotic particle decays or annihilations. We comment on possible scenarios behind neutron injection, however, we do not include a full treatment of the production and thermalization of neutrons in the code.}

\section{Implementation in the BBN code}
\addCorr
{
This code~\cite{Coc11} is based on the Big Bang model and $\Lambda$CDM cosmology. There are three pieces of evidence for this physical model: the universal expansion, cosmic microwave background (CMB) radiation, and BBN. The latter comes from the primordial abundances of the ``light cosmological elements'': \hli. They are produced during the first $\simeq$ 20 minutes of the Universe when it was dense and hot enough for nuclear reactions to take place. The number of free parameters entering the standard the BBN scenario has decreased with time. The number of light neutrino families is known from the measurement of the Z boson width by LEP experiments at CERN: $N_\nu = 2.9840 \pm 0.0082$~\cite{ALEPH:2005ab}. The lifetime of the neutron enters in weak reaction rate calculations and many nuclear reaction rates have been measured in nuclear physics laboratories. The last parameter to have been independently determined is the baryonic density of the Universe. It is now deduced from the observations of the anisotropies of the CMB radiation coming from the Wilkinson Microwave Anisotropy Probe (WMAP) satellite. The number of baryons per photon, which remains constant during the expansion, $\eta$ is directly related to $\Omega_b$ by $\Omega_bh^2 =  3. 65 \times 10^7\eta$. The WMAP 7 year results now give $\Omega_bh^2 =  0.02249 \pm 0.00056$ and $\eta =  (6.16 \pm 0.15)\times 10^{−10}$~\cite{Komatsu:2010fb}. In this context, primordial nucleosynthesis is a parameter-free theory and is the earliest probe of the Universe.
}

\addCorr
{
The main difficulty of the BBN calculations up to CNO is that an extensive network of reactions is needed, including $n$-, $p$-, $\alpha$-, but also $d$-, $t$-, and \tro-induced reactions. Most of the corresponding cross sections cannot be extracted from experimental data only. In the BBN code, we use a more reliable rate estimates provided by the TALYS reaction code~\cite{Goriely:2008zu}. 59 nuclides are included, from neutron to $^{23}$Na, linked by 391 reactions involving $n$-, $p$-, $d$-, $t$-, and \tro-induced reactions and 33 $\beta$-decay processes.}

Including an additional neutron injection in our SBBN code is straightforward. We allow protons to decay to neutrons with a lifetime of $\lambda^{-1}(t)$. As we are considering very low injection rates, this has no consequence on the high proton abundance. To illustrate the consequences of early or late injection, we consider the following cases:
\begin{enumerate}
\item $\lambda(t)$ = $\lambda_0$ at all time $t$ (or temperature $T$)
\item $\lambda(t)$ = $\lambda_0$ for $T\le T_c$ and 0 for $T>T_c$
\item $\lambda(t)$ = 0 for $T\le T_c$ and $\lambda_0$ for $T>T_c$
\item $\lambda(t)$ = $\lambda_0\exp(-t/\tau_x)$ 
\item $\lambda(t)$ = $\lambda_0 \left( \frac{T}{T_c}\right)^3$
\end{enumerate}
with $\lambda_0$ constant and $T_c$ = 0.2 and 0.3 GK. Since the proton abundance remains essentially constant ($Y_p\approx$0.5 to 0.7) during BBN the rate of injection $Y_p(t)\lambda(t)$ is constant in the (1), (2) and (3) intervals. Cases (4) and (5) represent more physical situations where neutrons come from the decay of a hypothetical particle $X$ of lifetime $\tau_x$ decaying to $X{\to}n+...$ with a branching ratio $B_n$, or as a product of the annihilation of dark matter particles (discuss 
\addCorr
{the relevant mass range of}
 $X$ is discussed below). In the latter cases the injection constant can be expressed as $\lambda_0=Y_x(t=0)B_n/Y_p\tau_x$.
\addCorr
{
A}
 cutoff in the neutron injection spectrum at lower temperatures (redshifts) is 
\addCorr{
expected
} due to the fact at a certain redshift (T $\sim$ 0.1-0.3 GK) the average time between interaction of neutrons becomes greater than the decay time of a neutron.

\section{Results}

\begin{figure}
\vskip -3cm
\includegraphics[width=\linewidth]{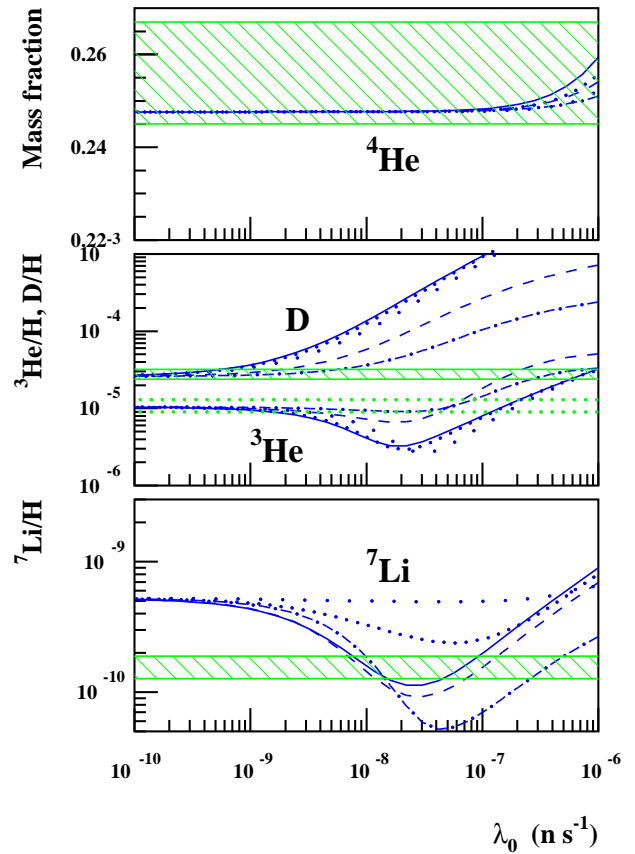}
\caption{\hli\ abundances as a function of neutron injection rate for cases (1) (solid line), (2) with $T_c$ = 0.2 (sparse dots) and 0.3 GK (dots) and case (3) with $T_c$ = 0.2 (dash) and 0.3 GK (dash-dot). Hatched zones represent the observational outcome (see the text for details).}
\label{f:heli}
\end{figure}

\begin{figure}
\includegraphics[width=\linewidth]{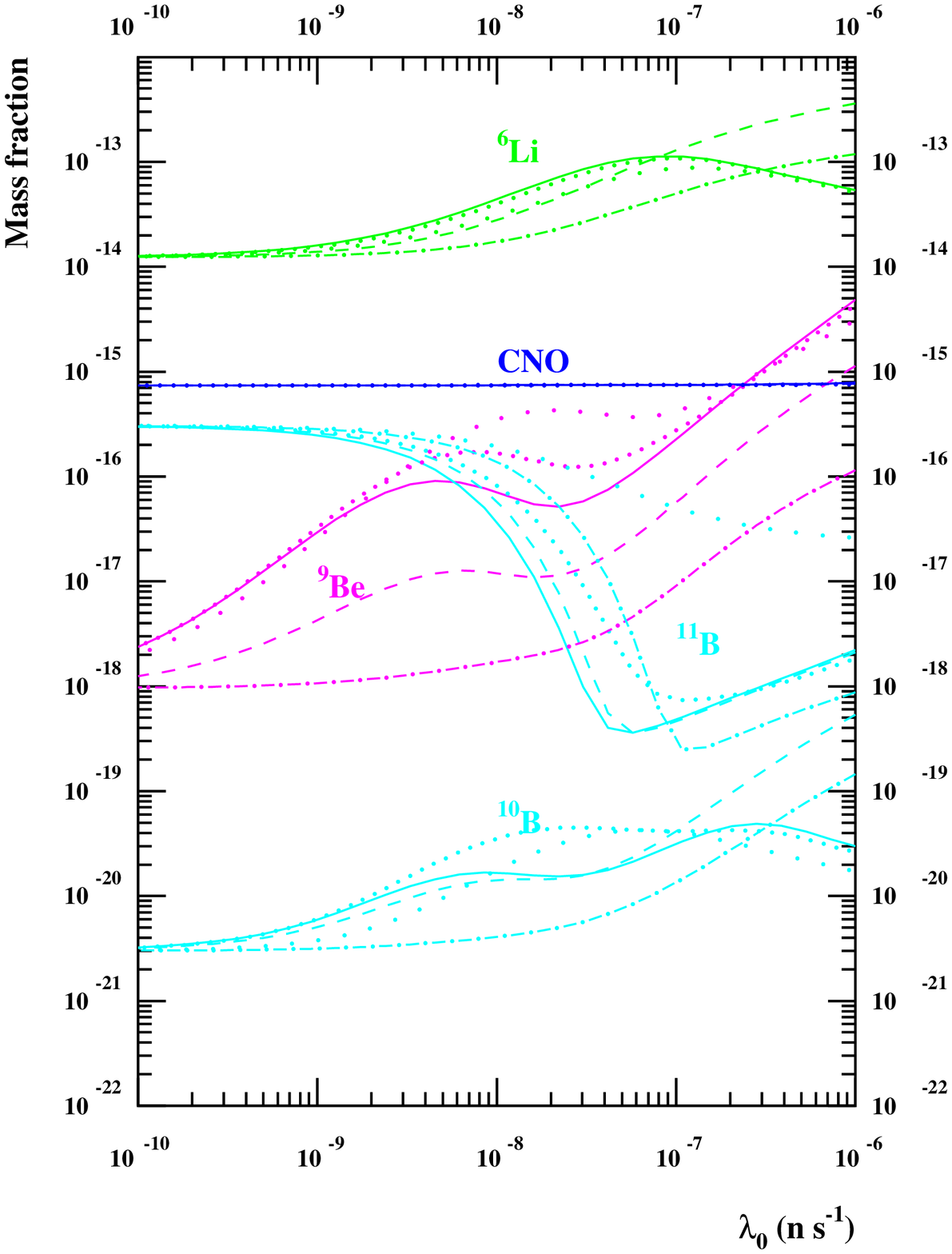}
\caption{Same as Fig.~\ref{f:heli} but for \six, \neu, \dix, \onz\ and CNO isotopes.}
\label{f:cno}
\end{figure}

\begin{figure}
\vskip -3cm
\includegraphics[width=\linewidth]{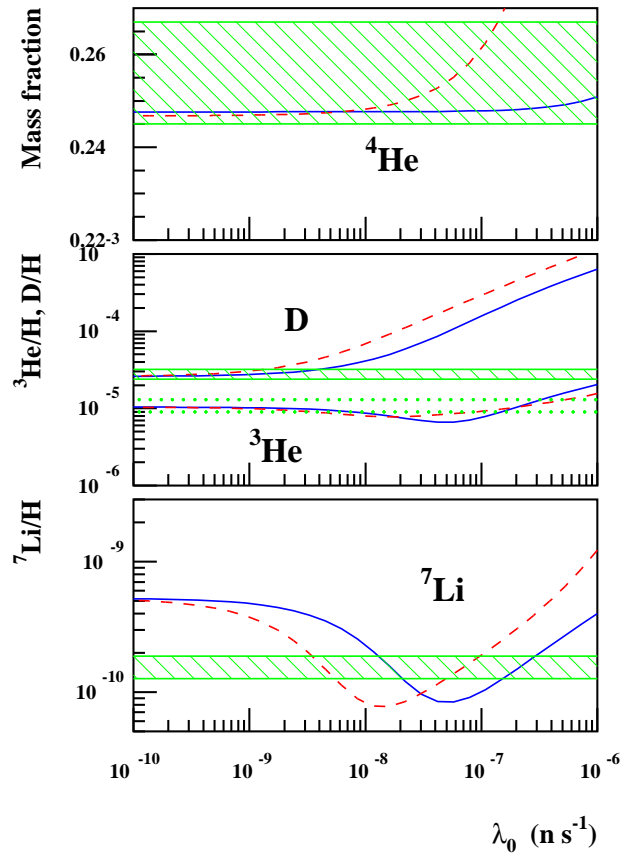}
\caption{\hli\ abundances as a function of neutron injection rate for case (4) i.e. decay, with $\tau_x$ = 40 mn (solid) and case (5) i.e. annihilation with $T_c$ = 0.3 GK (dash). Hatched zones represent the observational outcome (see the text for details).}
\label{f:heli1}
\end{figure}

\begin{figure}
\includegraphics[width=\linewidth]{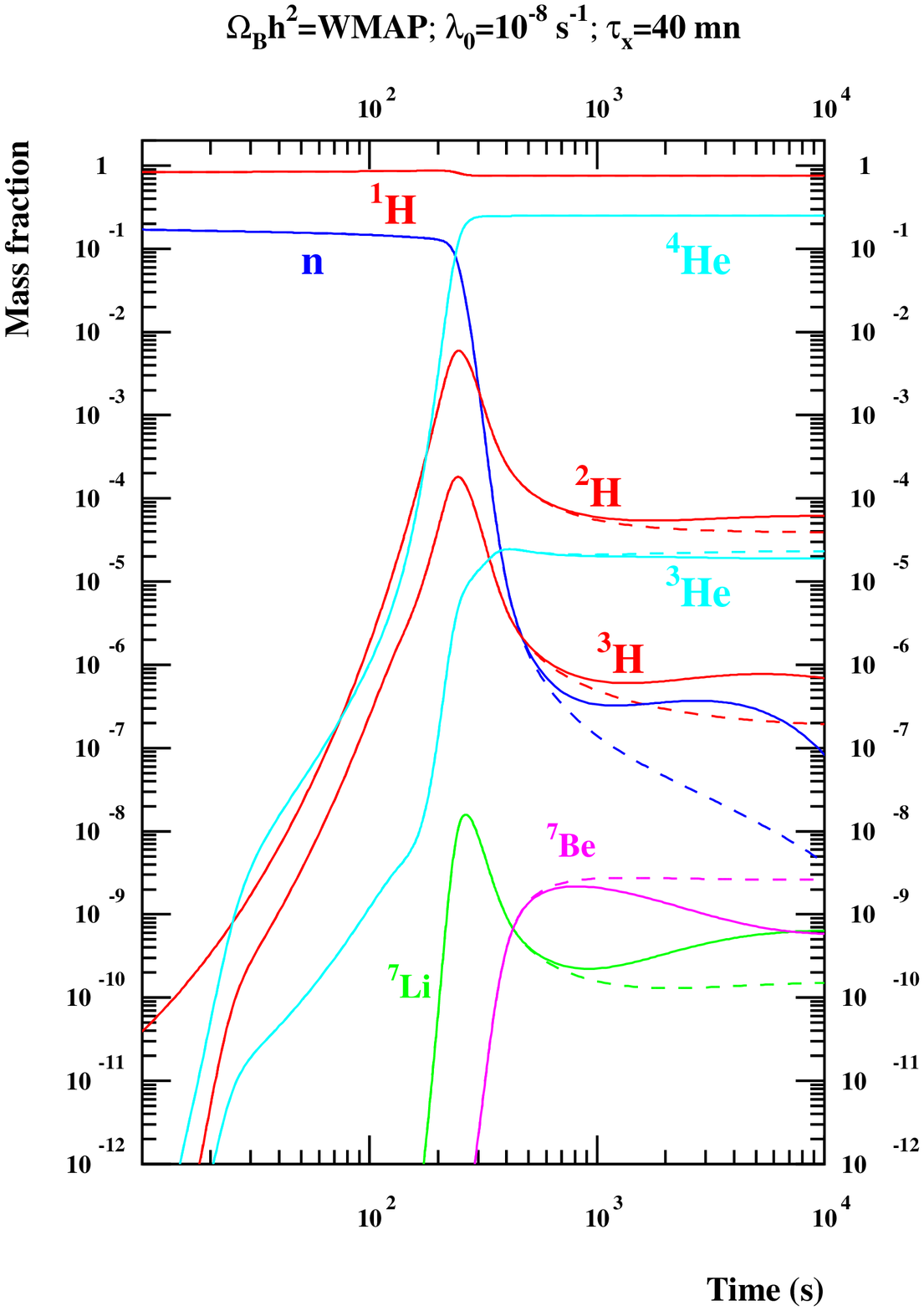}
\caption{\hli\ abundances in mass fraction as a function of time for $\lambda_0 = 0$ (dash) and case (4) with $\lambda_0=10^{-8}$~s$^{-1}$ and $\tau_x$ = 40 mn (solid). \addCorr{Note that, in this case, \sep\ is lowered at X(\sep)$ = 6 \times 10^{-10} $ which corresponds to Li/H$ = 1.1 \times 10^{-10} $.}}
\label{f:time}
\end{figure}

\addCorr{We summarize the outcome of the code in the figures.} Figure~\ref{f:heli} shows that injection of neutrons at a rate of $\lambda_0\approx10^{-8}$ s$^{-1}$ when $T$ $>$ 0.3~GK alleviates the \sep\ problem without significantly affecting the other isotopes. Figure~\ref{f:cno} shows that the \neu\ and \onz\ abundances depend strongly on $\lambda_0$ while the CNO abundances are not modified. There is a modest (less than 50\%) enhancement of \deu\, but this is well within astration uncertainties. As one might expect, exotic particle decays with $\lambda_0=(1.5-2)\times10^{-8}$ s$^{-1}$ or (1-30) GeV dark matter annihilations with $\lambda_0=(3-5)\times10^{-9}$ s$^{-1}$ (Figure~\ref{f:heli1}) help solve the \sep\ problem. \addCorr{Figure~\ref{f:time} shows the time evolution of the lightest elements for the SBBN and neutron injection from decaying exotic particle scenarios. The major impact of the injected neutrons is on \be\ and \sep\, helping diminish the primordial yield in \sep\, while the deviation on other light species show minor changes.}

\addCorr{In figures~\ref{f:heli} and~\ref{f:heli1} the different observational constraints are given (green hatched zones). As previously stated, the \sep/H abundance is obtained from~\cite{sbordone10}. The abundance of \deu\ is measured in quasar absorption systems. The weighted mean value of the observations is \deu/H=$(3.02\pm0.23)\times 10^{-5}$ (see~\cite{Olive:2012xf} for details). Note that two recent observations of \deu/H could slightly modify this value~\cite{Noterdaeme:2012pa} and~\cite{Pettini:2012ph}). Finally, the determination of the \qua\ abundance in extragalactic \textsc{H$_2$} regions is fraught with difficulties due to systematic errors. Consequently, as shown in figures~\ref{f:heli} and~\ref{f:heli1}, the weighted mean value is $Y_p=0.2566\pm0.0028$ still carries a large uncertainty~\cite{Aver:2010wq}.}

\section{Discussion}

While in our SBBN code, the neutrons are injected at equilibrium, it is likely that extra neutrons from any kind of beyond the Standard Model physics are produced out-of-equilibrium. It is, therefore, important to consider the thermalization process of neutrons during BBN before they decay, through which channels they do so, and to estimate the possible perturbations to SBBN abundances. As mentioned by \citet{Jedamzik:2004er,Jedamzik06}, the thermalization process calculation of energetic nucleons is simplified greatly by two facts: first, the Hubble time is much greater than the mean time between any of the interactions under consideration $\tau_H \approx 300 \left(\frac{T}{90 \, \mathrm{KeV}}\right)^{-2} \mathrm{s}$, and second, the interactions between non-thermal and thermal nucleons are unlikely. The decay time of free neutrons ($\tau_0 = 881 \,\mathrm{s}$) is even greater than the Hubble time.

\addCorr{There are} 3 main classes of reactions: (1) elastic and inelastic $\mathrm{n}-\mathrm{p}$ scattering, (2) the 
afore-mentioned spallation of ${}^4\mathrm{He}$ with production of ${}^3\mathrm{He}$, and (3) both elastic and inelastic scattering $\mathrm{n}-{}^4\mathrm{He}$. All of these processes contribute to thermalization, but the spallation to non-thermal ${}^3\mathrm{He}$ might disturb the abundance of ${}^6\mathrm{Li}$ \cite{Ellis11} through the following reactions
\begin{equation}
\begin{array}{lllllll}
n &+& {}^4\mathrm{He} &\to& {}^3\mathrm{He} &+& 2n \\
{}^4\mathrm{He} &+& {}^3\mathrm{He} &\to& {}^6\mathrm{Li} &+& p.
\end{array}
\end{equation}

Firstly, we justify the claim that the injected neutrons indeed thermalize before they decay and secondly, we estimate the production of ${}^3\mathrm{He}$ and ${}^6\mathrm{Li}$. 

\addCorr{\six\ is a very interesting isotope as a new cosmological nucleus. Indeed, its abundance is measured in low metallicity stars and offers a unique probe of two different mechanisms of nucleosynthesis: SBBN and cosmic rays. The former producing predominantly \sep, \six\ was until recently considered as a pure spallative --i.e., post BBN-- product. However, according to recent detections in very low metallicity PopII stars, the average is \six/\sep$\simeq 0.042$~\cite{Asplund:2005yt}. These observations have been interpreted as evidence for a large primitive abundance of \six\ (\six/H~$\simeq 10^{-12}$) while SBBN calculations confirm a low primordial value (\six/H~$\simeq 10^{-14}$). For details on the subject, see~\cite{Coc11,PhysRevC.82.065803}. Recently, new studies using 3D atmosphere model in metal poor halo stars reconsider the detection of \six. In~\cite{Steffen2012} two detections are confirmed (\six/\sep\ $\simeq5-10$\%). More observations are presently needed to improve the statistics. Nevertheless, this new spectroscopic research can be an indicator of new physics, as has been point out by~\citet{Jedamzik:1999di}. In that regard, we have to make sure that the extra physics we consider does not perturb significantly the abundance of \six. We do it in an order of magnitude estimate. Figure~\ref{f:cno} shows clearly that our models should not modify too much \six: at $\lambda_0\simeq10^{-8}$~n~s$^{-1}$ \six\ is in the range $(1-6)\times 10^{-14}$.}

The total cross-section of elastic and inelastic scatterings of n off p is about 70 mb at 100 MeV and 30-40 mb above 100 MeV. The cross-section of the reaction $\mathrm{n} + {}^4\mathrm{He} \to {}^3\mathrm{He} +2\mathrm{n}$ varies from about 15 mb at 30 MeV and 50 mb at 50 MeV down to 20 mb at 1 GeV. The sum of the cross-sections of the remaining inelastic processes is comparable to $\mathrm{n} + {}^4\mathrm{He} \to {}^3\mathrm{He} +2\mathrm{n}$ at 50 MeV and about 3-4 times greater than the rates above 100 MeV. The elastic scattering of neutrons off ${}^4\mathrm{He}$ varies from 500 mb at 20 MeV down to 30 mb at 100 MeV and stays constant at higher energies. 

The characteristic time of scattering of a neutron of kinetic energy $E_k$ off thermal H and ${}^4\mathrm{He}$ is 
\begin{align}
\tau_\sigma \approx A_i \!\left(\frac{T}{90\,\mathrm{KeV}}\right)^{-3}\! \left(\frac{\sigma}{50\,\mathrm{mb}}\right)^{-1} \nonumber \!\left(\frac{1\!+\!(m_n/50\,\mathrm{MeV})^2}{1+(m_n/E_k)^2}\right)^{-1/2} \mathrm{s}, 
\end{align}

where $A_H = 26$ and $A_{^4He} = 2.17$, $E_k$ is the kinetic energy of the neutron.
In the relativistic limit, $E_n \gg m_n$, $\tau$ goes down to 0.07 s, while the decay time grows linearly with energy $\tau_n = \frac{E}{m_n}\tau_0$. 

We also have to make sure that a significant fraction of energy is transferred in a single scattering or per mean free path length (its inverse is denoted $E (\lambda dE/dx)^{-1}$ in \citet{Jedamzik06}). In inelastic $\mathrm{n}-\mathrm{p}$ scatterings, $\frac{\Delta E}{E}$ remains constant up to 250 MeV ($\frac{\Delta E}{E} \gtrsim 0.1$). Indeed in elastic $\mathrm{n}-{}^4\mathrm{He}$ scatterings, the average energy transfer is about $5-10$ MeV. In inelastic processes $\mathrm{n}-{}^4\mathrm{He}$, the neutron loses as much as the binding energy of ${}^4\mathrm{He}$ (28.3 MeV) and a quarter of the remaining energy. 


From these arguments, it is clear that extra neutrons thermalize in these conditions before decaying in a range of temperatures from 100 KeV down to a few KeV and in the range of energies from about 10 MeV to 1 GeV.

The hypothesised extra neutrons might be produced by annihilating dark matter particles. The energy injection due to dark matter annihilations if the freeze-out of the dark matter species happened at BBN temperature is severely constrained (see~\cite{Sarkar:1995dd} for example). However, if the freeze-out happened before BBN, annihilations become marginal, as the expansion rate dominates the interaction rate. Nonetheless there would be a residual annihilation rate of dark matter into standard model particles. Eventually, after hadronization, a spectrum of neutrons would be generated, that would reach thermal equilibrium as discussed before.

The annihilation rate of uniformly distributed dark matter per baryon can be written as
\begin{equation}
\begin{array}{lll}
\Gamma_b &=&  \frac{1}{2}\langle \sigma v \rangle \frac{n^2_{0,X}}{n_{0,b}} \left(1+z\right)^3 \\
&=& 5.3\times 10^{-9}\left(\frac{\langle \sigma v \rangle}{3\times10^{-26} cm^3 s^{-1}}\right) \left(\frac{30 \, \mathrm{GeV}}{M_X}\right)^2 \left(\frac{T}{90\,\mathrm{keV}}\right)^3 s^{-1},
\end{array}
\end{equation}

where $n_{0,X}$ and $n_{0,b}$ are the present day number densities of dark matter and baryons. 
We see that at a temperature of about $90\;\mathrm{keV}$, a particle dark matter mass of $M_X \la 30 \,\mathrm{GeV}$ and a canonical annihilation rate are plausible parameters needed to achieve $\lambda = (3-5) \times 10^{-9}$~s$^{-1}$, depending on the neutron spectrum generated by the annihilations.

The neutron spectrum is generated after hadronization of the particles produced at annihilation, and it is expected to be peaked at roughly $M_X/5 - M_X/15$. Therefore, the lighter the dark matter particle, the larger the fraction of thermal neutrons. However, the dark matter has to be heavy enough to produce neutrons, hence, the most interesting mass range lies roughly between 1 and 30 GeV.

The annihilation rate and branching ratios depend on the dark matter candidate. Moreover, the dark matter temperature evolves from chemical decoupling down to thermal decoupling (see~\cite{Bringmann:2006mu}). The dependence of the annihilation rate on the dark matter temperature can be very strong; for example, if the freeze-out mechanism invokes a nearly resonant exchange, or co-annihilations~\cite{Griest:1990kh}.

A relevant example for a dark matter candidate in the mass range discussed here is the neutralino in the Next-to-Minimal Supersymmetric Standard Model. As shown in~\cite{AlbornozVasquez:2011js}, the resonant mechanism at freeze-out can yield a very large boost to the annihilation rate at lower temperatures (see Fig. 4 in~\cite{AlbornozVasquez:2011js}). For kinetic decoupling at $T_{kd}\sim T_{fo}/10$, one could have a factor $10-100$ enhancement in the annihilation rate from the $3\times10^{-26}\rm{cm}^3\rm{s}^{-1}$ required at freeze-out. Thus there can be a variety of dark matter candidates (with different masses and annihilation cross-section mechanisms) which provide an injection flux of $(3-5) \times 10^{-9}$~s$^{-1}$.

\addCorr{It is interesting to note that some of these candidates could explain direct detection signals as they have the right mass range and could attain the needed interaction rates with detectors. Also, they could be challenged by $\gamma$-ray production at dwarf spheroidal galaxies. Relating annihilating rates at freeze-out, BBN and galactic times, and elastic scattering interactions with nuclei, can provide powerful constraints on a given dark matter model.}


In conclusion, neutron injection can help to resolve the \sep\ problem provided that the neutrons are essentially thermal. This can be achieved for annihilations or decays of dark matter particles in the mass range 1-30 GeV. A detailed physical model involving, for example, a metastable supersymmetric NLSP or annihilating neutralino dark matter is beyond the scope of this paper, but would seem to be easily achievable.

\section*{Acknowledgements}
This work was sponsored by the ``VACOUL'' Program sposored by the French Agence Nationale pour la Recherche (ANR) and by the European Research Council (ERC) Advanced Grant ``Dark Matters (DARK)''. 

\bibliography{bbn_n10}

\begin{thebibliography}{42}
\expandafter\ifx\csname natexlab\endcsname\relax\def\natexlab#1{#1}\fi
\expandafter\ifx\csname bibnamefont\endcsname\relax
  \def\bibnamefont#1{#1}\fi
\expandafter\ifx\csname bibfnamefont\endcsname\relax
  \def\bibfnamefont#1{#1}\fi
\expandafter\ifx\csname citenamefont\endcsname\relax
  \def\citenamefont#1{#1}\fi
\expandafter\ifx\csname url\endcsname\relax
  \def\url#1{\texttt{#1}}\fi
\expandafter\ifx\csname urlprefix\endcsname\relax\def\urlprefix{URL }\fi
\providecommand{\bibinfo}[2]{#2}
\providecommand{\eprint}[2][]{\url{#2}}

\bibitem[{\citenamefont{{Spite} and {Spite}}(1982)}]{Spite82}
\bibinfo{author}{\bibfnamefont{F.}~\bibnamefont{{Spite}}} \bibnamefont{and}
  \bibinfo{author}{\bibfnamefont{M.}~\bibnamefont{{Spite}}},
  \bibinfo{journal}{A \& A} \textbf{\bibinfo{volume}{115}},
  \bibinfo{pages}{357} (\bibinfo{year}{1982}).

\bibitem[{\citenamefont{Ryan et~al.}(2000)\citenamefont{Ryan, Beers, Olive,
  Fields, and Norris}}]{Ryanetal2000}
\bibinfo{author}{\bibfnamefont{S.~G.} \bibnamefont{Ryan}},
  \bibinfo{author}{\bibfnamefont{T.~C.} \bibnamefont{Beers}},
  \bibinfo{author}{\bibfnamefont{K.~A.} \bibnamefont{Olive}},
  \bibinfo{author}{\bibfnamefont{B.~D.} \bibnamefont{Fields}},
  \bibnamefont{and} \bibinfo{author}{\bibfnamefont{J.~E.}
  \bibnamefont{Norris}}, \bibinfo{journal}{Astrophys.J.}
  \textbf{\bibinfo{volume}{530}}, \bibinfo{pages}{L57} (\bibinfo{year}{2000}).

\bibitem[{\citenamefont{{Sbordone} et~al.}(2010)\citenamefont{{Sbordone},
  {Bonifacio}, {Caffau}, {Ludwig}, {Behara}, {Gonz{\'a}lez Hern{\'a}ndez},
  {Steffen}, {Cayrel}, {Freytag}, {van't Veer} et~al.}}]{sbordone10}
\bibinfo{author}{\bibfnamefont{L.}~\bibnamefont{{Sbordone}}},
  \bibinfo{author}{\bibfnamefont{P.}~\bibnamefont{{Bonifacio}}},
  \bibinfo{author}{\bibfnamefont{E.}~\bibnamefont{{Caffau}}},
  \bibinfo{author}{\bibfnamefont{H.-G.} \bibnamefont{{Ludwig}}},
  \bibinfo{author}{\bibfnamefont{N.~T.} \bibnamefont{{Behara}}},
  \bibinfo{author}{\bibfnamefont{J.~I.} \bibnamefont{{Gonz{\'a}lez
  Hern{\'a}ndez}}},
  \bibinfo{author}{\bibfnamefont{M.}~\bibnamefont{{Steffen}}},
  \bibinfo{author}{\bibfnamefont{R.}~\bibnamefont{{Cayrel}}},
  \bibinfo{author}{\bibfnamefont{B.}~\bibnamefont{{Freytag}}},
  \bibinfo{author}{\bibfnamefont{C.}~\bibnamefont{{van't Veer}}},
  \bibnamefont{et~al.}, \bibinfo{journal}{A \& A}
  \textbf{\bibinfo{volume}{522}}, \bibinfo{eid}{A26} (\bibinfo{year}{2010}),
  \eprint{1003.4510}.

\bibitem[{\citenamefont{{Spite} and {Spite}}(2010)}]{Spite10}
\bibinfo{author}{\bibfnamefont{M.}~\bibnamefont{{Spite}}} \bibnamefont{and}
  \bibinfo{author}{\bibfnamefont{F.}~\bibnamefont{{Spite}}}, in
  \emph{\bibinfo{booktitle}{IAU Symposium}}, edited by
  \bibinfo{editor}{\bibnamefont{{C.~Charbonnel, M.~Tosi, F.~Primas, \&
  C.~Chiappini}}} (\bibinfo{year}{2010}), vol. \bibinfo{volume}{268} of
  \emph{\bibinfo{series}{IAU Symposium}}, pp. \bibinfo{pages}{201--210},
  \eprint{1002.1004}.

\bibitem[{\citenamefont{Frebel and Norris}(2011)}]{frebel11}
\bibinfo{author}{\bibfnamefont{A.}~\bibnamefont{Frebel}} \bibnamefont{and}
  \bibinfo{author}{\bibfnamefont{J.~E.} \bibnamefont{Norris}}
  (\bibinfo{year}{2011}), \eprint{1102.1748}.

\bibitem[{\citenamefont{Coc and Vangioni}(2010)}]{CV10}
\bibinfo{author}{\bibfnamefont{A.}~\bibnamefont{Coc}} \bibnamefont{and}
  \bibinfo{author}{\bibfnamefont{E.}~\bibnamefont{Vangioni}},
  \bibinfo{journal}{J.Phys.Conf.Ser.} \textbf{\bibinfo{volume}{202}},
  \bibinfo{pages}{012001} (\bibinfo{year}{2010}).

\bibitem[{\citenamefont{Coc et~al.}(2004)\citenamefont{Coc, Vangioni-Flam,
  Descouvemont, Adahchour, and Angulo}}]{Coc04}
\bibinfo{author}{\bibfnamefont{A.}~\bibnamefont{Coc}},
  \bibinfo{author}{\bibfnamefont{E.}~\bibnamefont{Vangioni-Flam}},
  \bibinfo{author}{\bibfnamefont{P.}~\bibnamefont{Descouvemont}},
  \bibinfo{author}{\bibfnamefont{A.}~\bibnamefont{Adahchour}},
  \bibnamefont{and} \bibinfo{author}{\bibfnamefont{C.}~\bibnamefont{Angulo}},
  \bibinfo{journal}{Astrophys.J.} \textbf{\bibinfo{volume}{600}},
  \bibinfo{pages}{544} (\bibinfo{year}{2004}), \eprint{astro-ph/0309480}.

\bibitem[{\citenamefont{Cyburt and Pospelov}(2009)}]{Cyb09}
\bibinfo{author}{\bibfnamefont{R.~H.} \bibnamefont{Cyburt}} \bibnamefont{and}
  \bibinfo{author}{\bibfnamefont{M.}~\bibnamefont{Pospelov}}
  (\bibinfo{year}{2009}), \eprint{0906.4373}.

\bibitem[{\citenamefont{Angulo et~al.}(2005)\citenamefont{Angulo, Casarejos,
  Couder, Demaret, Leleux et~al.}}]{Ang05}
\bibinfo{author}{\bibfnamefont{C.}~\bibnamefont{Angulo}},
  \bibinfo{author}{\bibfnamefont{E.}~\bibnamefont{Casarejos}},
  \bibinfo{author}{\bibfnamefont{M.}~\bibnamefont{Couder}},
  \bibinfo{author}{\bibfnamefont{P.}~\bibnamefont{Demaret}},
  \bibinfo{author}{\bibfnamefont{P.}~\bibnamefont{Leleux}},
  \bibnamefont{et~al.}, \bibinfo{journal}{Astrophys.J.}
  \textbf{\bibinfo{volume}{630}}, \bibinfo{pages}{L105} (\bibinfo{year}{2005}),
  \eprint{astro-ph/0508454}.

\bibitem[{\citenamefont{O'Malley et~al.}(2011)\citenamefont{O'Malley, Bardayan,
  Adekola, Ahn, Chae et~al.}}]{OMa11}
\bibinfo{author}{\bibfnamefont{P.}~\bibnamefont{O'Malley}},
  \bibinfo{author}{\bibfnamefont{D.}~\bibnamefont{Bardayan}},
  \bibinfo{author}{\bibfnamefont{A.}~\bibnamefont{Adekola}},
  \bibinfo{author}{\bibfnamefont{S.}~\bibnamefont{Ahn}},
  \bibinfo{author}{\bibfnamefont{K.}~\bibnamefont{Chae}}, \bibnamefont{et~al.},
  \bibinfo{journal}{Phys.Rev.} \textbf{\bibinfo{volume}{C84}},
  \bibinfo{pages}{042801} (\bibinfo{year}{2011}).

\bibitem[{\citenamefont{Kirsebom and Davids}(2011)}]{Kir11}
\bibinfo{author}{\bibfnamefont{O.}~\bibnamefont{Kirsebom}} \bibnamefont{and}
  \bibinfo{author}{\bibfnamefont{B.}~\bibnamefont{Davids}},
  \bibinfo{journal}{Phys.Rev.} \textbf{\bibinfo{volume}{C84}},
  \bibinfo{pages}{058801} (\bibinfo{year}{2011}), \eprint{1109.4690}.

\bibitem[{\citenamefont{Chakraborty et~al.}(2011)\citenamefont{Chakraborty,
  Fields, and Olive}}]{Cha11}
\bibinfo{author}{\bibfnamefont{N.}~\bibnamefont{Chakraborty}},
  \bibinfo{author}{\bibfnamefont{B.~D.} \bibnamefont{Fields}},
  \bibnamefont{and} \bibinfo{author}{\bibfnamefont{K.~A.} \bibnamefont{Olive}},
  \bibinfo{journal}{Phys.Rev.} \textbf{\bibinfo{volume}{D83}},
  \bibinfo{pages}{063006} (\bibinfo{year}{2011}), \eprint{1011.0722}.

\bibitem[{\citenamefont{Reno and Seckel}(1988)}]{Reno:1987qw}
\bibinfo{author}{\bibfnamefont{M.}~\bibnamefont{Reno}} \bibnamefont{and}
  \bibinfo{author}{\bibfnamefont{D.}~\bibnamefont{Seckel}},
  \bibinfo{journal}{Phys.Rev.} \textbf{\bibinfo{volume}{D37}},
  \bibinfo{pages}{3441} (\bibinfo{year}{1988}).

\bibitem[{\citenamefont{Coc et~al.}(2007)\citenamefont{Coc, Nunes, Olive, Uzan,
  and Vangioni}}]{Coc07}
\bibinfo{author}{\bibfnamefont{A.}~\bibnamefont{Coc}},
  \bibinfo{author}{\bibfnamefont{N.~J.} \bibnamefont{Nunes}},
  \bibinfo{author}{\bibfnamefont{K.~A.} \bibnamefont{Olive}},
  \bibinfo{author}{\bibfnamefont{J.-P.} \bibnamefont{Uzan}}, \bibnamefont{and}
  \bibinfo{author}{\bibfnamefont{E.}~\bibnamefont{Vangioni}},
  \bibinfo{journal}{Phys.Rev.} \textbf{\bibinfo{volume}{D76}},
  \bibinfo{pages}{023511} (\bibinfo{year}{2007}), \eprint{astro-ph/0610733}.

\bibitem[{\citenamefont{Coc et~al.}(2012)\citenamefont{Coc, Goriely, Xu,
  Saimpert, and Vangioni}}]{Coc11}
\bibinfo{author}{\bibfnamefont{A.}~\bibnamefont{Coc}},
  \bibinfo{author}{\bibfnamefont{S.}~\bibnamefont{Goriely}},
  \bibinfo{author}{\bibfnamefont{Y.}~\bibnamefont{Xu}},
  \bibinfo{author}{\bibfnamefont{M.}~\bibnamefont{Saimpert}}, \bibnamefont{and}
  \bibinfo{author}{\bibfnamefont{E.}~\bibnamefont{Vangioni}},
  \bibinfo{journal}{Astrophys.J.} \textbf{\bibinfo{volume}{744}},
  \bibinfo{pages}{158} (\bibinfo{year}{2012}), \eprint{1107.1117}.

\bibitem[{\citenamefont{Cyburt et~al.}(2008)\citenamefont{Cyburt, Fields, and
  Olive}}]{Cyb08}
\bibinfo{author}{\bibfnamefont{R.~H.} \bibnamefont{Cyburt}},
  \bibinfo{author}{\bibfnamefont{B.~D.} \bibnamefont{Fields}},
  \bibnamefont{and} \bibinfo{author}{\bibfnamefont{K.~A.} \bibnamefont{Olive}},
  \bibinfo{journal}{JCAP} \textbf{\bibinfo{volume}{0811}}, \bibinfo{pages}{012}
  (\bibinfo{year}{2008}), \eprint{0808.2818}.

\bibitem[{\citenamefont{Malaney and Mathews}(1993)}]{Malaney:1993ah}
\bibinfo{author}{\bibfnamefont{R.}~\bibnamefont{Malaney}} \bibnamefont{and}
  \bibinfo{author}{\bibfnamefont{G.}~\bibnamefont{Mathews}},
  \bibinfo{journal}{Phys.Rept.} \textbf{\bibinfo{volume}{229}},
  \bibinfo{pages}{145} (\bibinfo{year}{1993}).

\bibitem[{\citenamefont{Sarkar}(1996)}]{Sarkar:1995dd}
\bibinfo{author}{\bibfnamefont{S.}~\bibnamefont{Sarkar}},
  \bibinfo{journal}{Rept.Prog.Phys.} \textbf{\bibinfo{volume}{59}},
  \bibinfo{pages}{1493} (\bibinfo{year}{1996}), \bibinfo{note}{dedicated to
  Dennis Sciama on his 67th birthday}, \eprint{hep-ph/9602260}.

\bibitem[{\citenamefont{Jedamzik and Pospelov}(2009)}]{Jedamzik:2009uy}
\bibinfo{author}{\bibfnamefont{K.}~\bibnamefont{Jedamzik}} \bibnamefont{and}
  \bibinfo{author}{\bibfnamefont{M.}~\bibnamefont{Pospelov}},
  \bibinfo{journal}{New J.Phys.} \textbf{\bibinfo{volume}{11}},
  \bibinfo{pages}{105028} (\bibinfo{year}{2009}), \bibinfo{note}{a mini-review
  for the NJP special issue on dark matter}, \eprint{0906.2087}.

\bibitem[{\citenamefont{Pospelov and
  Pradler}(2010{\natexlab{a}})}]{Pospelov:2010hj}
\bibinfo{author}{\bibfnamefont{M.}~\bibnamefont{Pospelov}} \bibnamefont{and}
  \bibinfo{author}{\bibfnamefont{J.}~\bibnamefont{Pradler}},
  \bibinfo{journal}{Ann.Rev.Nucl.Part.Sci.} \textbf{\bibinfo{volume}{60}},
  \bibinfo{pages}{539} (\bibinfo{year}{2010}{\natexlab{a}}),
  \eprint{1011.1054}.

\bibitem[{\citenamefont{Kohri and Santoso}(2009)}]{PhysRevD.79.043514}
\bibinfo{author}{\bibfnamefont{K.}~\bibnamefont{Kohri}} \bibnamefont{and}
  \bibinfo{author}{\bibfnamefont{Y.}~\bibnamefont{Santoso}},
  \bibinfo{journal}{Phys. Rev. D} \textbf{\bibinfo{volume}{79}},
  \bibinfo{pages}{043514} (\bibinfo{year}{2009}).

\bibitem[{\citenamefont{{Dimopoulos} et~al.}(1988)\citenamefont{{Dimopoulos},
  {Esmailzadeh}, {Starkman}, and {Hall}}}]{1988ApJ...330..545D}
\bibinfo{author}{\bibfnamefont{S.}~\bibnamefont{{Dimopoulos}}},
  \bibinfo{author}{\bibfnamefont{R.}~\bibnamefont{{Esmailzadeh}}},
  \bibinfo{author}{\bibfnamefont{G.~D.} \bibnamefont{{Starkman}}},
  \bibnamefont{and} \bibinfo{author}{\bibfnamefont{L.~J.}
  \bibnamefont{{Hall}}}, \bibinfo{journal}{\apj}
  \textbf{\bibinfo{volume}{330}}, \bibinfo{pages}{545} (\bibinfo{year}{1988}).

\bibitem[{\citenamefont{Jedamzik}(2004)}]{Jedamzik:2004er}
\bibinfo{author}{\bibfnamefont{K.}~\bibnamefont{Jedamzik}},
  \bibinfo{journal}{Phys.Rev.} \textbf{\bibinfo{volume}{D70}},
  \bibinfo{pages}{063524} (\bibinfo{year}{2004}), \eprint{astro-ph/0402344}.

\bibitem[{\citenamefont{Jedamzik}(2006)}]{Jedamzik06}
\bibinfo{author}{\bibfnamefont{K.}~\bibnamefont{Jedamzik}},
  \bibinfo{journal}{Phys.Rev.} \textbf{\bibinfo{volume}{D74}},
  \bibinfo{pages}{103509} (\bibinfo{year}{2006}), \eprint{hep-ph/0604251}.

\bibitem[{\citenamefont{Cumberbatch et~al.}(2007)\citenamefont{Cumberbatch,
  Ichikawa, Kawasaki, Kohri, Silk et~al.}}]{Cumberbatch:2007me}
\bibinfo{author}{\bibfnamefont{D.}~\bibnamefont{Cumberbatch}},
  \bibinfo{author}{\bibfnamefont{K.}~\bibnamefont{Ichikawa}},
  \bibinfo{author}{\bibfnamefont{M.}~\bibnamefont{Kawasaki}},
  \bibinfo{author}{\bibfnamefont{K.}~\bibnamefont{Kohri}},
  \bibinfo{author}{\bibfnamefont{J.}~\bibnamefont{Silk}}, \bibnamefont{et~al.},
  \bibinfo{journal}{Phys.Rev.} \textbf{\bibinfo{volume}{D76}},
  \bibinfo{pages}{123005} (\bibinfo{year}{2007}), \eprint{0708.0095}.

\bibitem[{\citenamefont{Cyburt et~al.}(2010)\citenamefont{Cyburt, Ellis,
  Fields, Luo, Olive et~al.}}]{Cyburt:2010vz}
\bibinfo{author}{\bibfnamefont{R.~H.} \bibnamefont{Cyburt}},
  \bibinfo{author}{\bibfnamefont{J.}~\bibnamefont{Ellis}},
  \bibinfo{author}{\bibfnamefont{B.~D.} \bibnamefont{Fields}},
  \bibinfo{author}{\bibfnamefont{F.}~\bibnamefont{Luo}},
  \bibinfo{author}{\bibfnamefont{K.~A.} \bibnamefont{Olive}},
  \bibnamefont{et~al.}, \bibinfo{journal}{JCAP}
  \textbf{\bibinfo{volume}{1010}}, \bibinfo{pages}{032} (\bibinfo{year}{2010}),
  \eprint{1007.4173}.

\bibitem[{\citenamefont{Pospelov and
  Pradler}(2010{\natexlab{b}})}]{Pospelov:2010cw}
\bibinfo{author}{\bibfnamefont{M.}~\bibnamefont{Pospelov}} \bibnamefont{and}
  \bibinfo{author}{\bibfnamefont{J.}~\bibnamefont{Pradler}},
  \bibinfo{journal}{Phys.Rev.} \textbf{\bibinfo{volume}{D82}},
  \bibinfo{pages}{103514} (\bibinfo{year}{2010}{\natexlab{b}}),
  \eprint{1006.4172}.

\bibitem[{691576()}]{ALEPH:2005ab}
691576, \bibinfo{journal}{Phys.Rept.} \textbf{\bibinfo{volume}{427}},
  \bibinfo{pages}{257} (\bibinfo{year}{2006}), \eprint{hep-ex/0509008}.

\bibitem[{\citenamefont{Komatsu et~al.}(2011)}]{Komatsu:2010fb}
\bibinfo{author}{\bibfnamefont{E.}~\bibnamefont{Komatsu}} \bibnamefont{et~al.}
  (\bibinfo{collaboration}{WMAP Collaboration}),
  \bibinfo{journal}{Astrophys.J.Suppl.} \textbf{\bibinfo{volume}{192}},
  \bibinfo{pages}{18} (\bibinfo{year}{2011}), \eprint{1001.4538}.

\bibitem[{\citenamefont{Goriely et~al.}(2008)\citenamefont{Goriely, Hilaire,
  and Koning}}]{Goriely:2008zu}
\bibinfo{author}{\bibfnamefont{S.}~\bibnamefont{Goriely}},
  \bibinfo{author}{\bibfnamefont{S.}~\bibnamefont{Hilaire}}, \bibnamefont{and}
  \bibinfo{author}{\bibfnamefont{A.}~\bibnamefont{Koning}},
  \bibinfo{journal}{Astron.Astrophys.} \textbf{\bibinfo{volume}{487}},
  \bibinfo{pages}{767} (\bibinfo{year}{2008}), \eprint{0806.2239}.

\bibitem[{\citenamefont{Olive et~al.}(2012)\citenamefont{Olive, Petitjean,
  Vangioni, and Silk}}]{Olive:2012xf}
\bibinfo{author}{\bibfnamefont{K.~A.} \bibnamefont{Olive}},
  \bibinfo{author}{\bibfnamefont{P.}~\bibnamefont{Petitjean}},
  \bibinfo{author}{\bibfnamefont{E.}~\bibnamefont{Vangioni}}, \bibnamefont{and}
  \bibinfo{author}{\bibfnamefont{J.}~\bibnamefont{Silk}},
  \bibinfo{journal}{submitted to \mnras}  (\bibinfo{year}{2012}),
  \eprint{1203.5701}.

\bibitem[{\citenamefont{Noterdaeme et~al.}(2012)\citenamefont{Noterdaeme,
  Lopez, Dumont, Ledoux, Molaro et~al.}}]{Noterdaeme:2012pa}
\bibinfo{author}{\bibfnamefont{P.}~\bibnamefont{Noterdaeme}},
  \bibinfo{author}{\bibfnamefont{S.}~\bibnamefont{Lopez}},
  \bibinfo{author}{\bibfnamefont{V.}~\bibnamefont{Dumont}},
  \bibinfo{author}{\bibfnamefont{C.}~\bibnamefont{Ledoux}},
  \bibinfo{author}{\bibfnamefont{P.}~\bibnamefont{Molaro}},
  \bibnamefont{et~al.}, \bibinfo{journal}{accepted in \aap}
  (\bibinfo{year}{2012}), \eprint{1205.3777}.

\bibitem[{\citenamefont{Pettini and Cooke}(2012)}]{Pettini:2012ph}
\bibinfo{author}{\bibfnamefont{M.}~\bibnamefont{Pettini}} \bibnamefont{and}
  \bibinfo{author}{\bibfnamefont{R.}~\bibnamefont{Cooke}}
  (\bibinfo{year}{2012}), \eprint{1205.3785}.

\bibitem[{\citenamefont{Aver et~al.}(2010)\citenamefont{Aver, Olive, and
  Skillman}}]{Aver:2010wq}
\bibinfo{author}{\bibfnamefont{E.}~\bibnamefont{Aver}},
  \bibinfo{author}{\bibfnamefont{K.~A.} \bibnamefont{Olive}}, \bibnamefont{and}
  \bibinfo{author}{\bibfnamefont{E.~D.} \bibnamefont{Skillman}},
  \bibinfo{journal}{JCAP} \textbf{\bibinfo{volume}{1005}}, \bibinfo{pages}{003}
  (\bibinfo{year}{2010}), \eprint{1001.5218}.

\bibitem[{\citenamefont{Ellis et~al.}(2011)\citenamefont{Ellis, Fields, Luo,
  Olive, and Spanos}}]{Ellis11}
\bibinfo{author}{\bibfnamefont{J.}~\bibnamefont{Ellis}},
  \bibinfo{author}{\bibfnamefont{B.~D.} \bibnamefont{Fields}},
  \bibinfo{author}{\bibfnamefont{F.}~\bibnamefont{Luo}},
  \bibinfo{author}{\bibfnamefont{K.~A.} \bibnamefont{Olive}}, \bibnamefont{and}
  \bibinfo{author}{\bibfnamefont{V.~C.} \bibnamefont{Spanos}},
  \bibinfo{journal}{Phys.Rev.} \textbf{\bibinfo{volume}{D84}},
  \bibinfo{pages}{123502} (\bibinfo{year}{2011}), \eprint{1109.0549}.

\bibitem[{\citenamefont{Asplund et~al.}(2006)\citenamefont{Asplund, Lambert,
  Nissen, Primas, and Smith}}]{Asplund:2005yt}
\bibinfo{author}{\bibfnamefont{M.}~\bibnamefont{Asplund}},
  \bibinfo{author}{\bibfnamefont{D.~L.} \bibnamefont{Lambert}},
  \bibinfo{author}{\bibfnamefont{P.~E.} \bibnamefont{Nissen}},
  \bibinfo{author}{\bibfnamefont{F.}~\bibnamefont{Primas}}, \bibnamefont{and}
  \bibinfo{author}{\bibfnamefont{V.~V.} \bibnamefont{Smith}},
  \bibinfo{journal}{Astrophys.J.} \textbf{\bibinfo{volume}{644}},
  \bibinfo{pages}{229} (\bibinfo{year}{2006}), \eprint{astro-ph/0510636}.

\bibitem[{\citenamefont{Hammache et~al.}(2010)\citenamefont{Hammache, Heil,
  Typel, Galaviz, S\"ummerer, Coc, Uhlig, Attallah, Caamano, Cortina
  et~al.}}]{PhysRevC.82.065803}
\bibinfo{author}{\bibfnamefont{F.}~\bibnamefont{Hammache}},
  \bibinfo{author}{\bibfnamefont{M.}~\bibnamefont{Heil}},
  \bibinfo{author}{\bibfnamefont{S.}~\bibnamefont{Typel}},
  \bibinfo{author}{\bibfnamefont{D.}~\bibnamefont{Galaviz}},
  \bibinfo{author}{\bibfnamefont{K.}~\bibnamefont{S\"ummerer}},
  \bibinfo{author}{\bibfnamefont{A.}~\bibnamefont{Coc}},
  \bibinfo{author}{\bibfnamefont{F.}~\bibnamefont{Uhlig}},
  \bibinfo{author}{\bibfnamefont{F.}~\bibnamefont{Attallah}},
  \bibinfo{author}{\bibfnamefont{M.}~\bibnamefont{Caamano}},
  \bibinfo{author}{\bibfnamefont{D.}~\bibnamefont{Cortina}},
  \bibnamefont{et~al.}, \bibinfo{journal}{Phys. Rev. C}
  \textbf{\bibinfo{volume}{82}}, \bibinfo{pages}{065803}
  (\bibinfo{year}{2010}).

\bibitem[{\citenamefont{Steffen et~al.}(2012)}]{Steffen2012}
\bibinfo{author}{\bibfnamefont{M.}~\bibnamefont{Steffen}} \bibnamefont{et~al.},
  \bibinfo{journal}{Mem.S.A.It.Suppl.} \textbf{\bibinfo{volume}{22}},
  \bibinfo{pages}{135} (\bibinfo{year}{2012}).

\bibitem[{\citenamefont{Jedamzik}(2000)}]{Jedamzik:1999di}
\bibinfo{author}{\bibfnamefont{K.}~\bibnamefont{Jedamzik}},
  \bibinfo{journal}{Phys.Rev.Lett.} \textbf{\bibinfo{volume}{84}},
  \bibinfo{pages}{3248} (\bibinfo{year}{2000}), \eprint{astro-ph/9909445}.

\bibitem[{\citenamefont{Bringmann and Hofmann}(2007)}]{Bringmann:2006mu}
\bibinfo{author}{\bibfnamefont{T.}~\bibnamefont{Bringmann}} \bibnamefont{and}
  \bibinfo{author}{\bibfnamefont{S.}~\bibnamefont{Hofmann}},
  \bibinfo{journal}{JCAP} \textbf{\bibinfo{volume}{0407}}, \bibinfo{pages}{016}
  (\bibinfo{year}{2007}), \eprint{hep-ph/0612238}.

\bibitem[{\citenamefont{Griest and Seckel}(1991)}]{Griest:1990kh}
\bibinfo{author}{\bibfnamefont{K.}~\bibnamefont{Griest}} \bibnamefont{and}
  \bibinfo{author}{\bibfnamefont{D.}~\bibnamefont{Seckel}},
  \bibinfo{journal}{Phys.Rev.} \textbf{\bibinfo{volume}{D43}},
  \bibinfo{pages}{3191} (\bibinfo{year}{1991}).

\bibitem[{\citenamefont{Albornoz~V\'asquez
  et~al.}(2011)\citenamefont{Albornoz~V\'asquez, B\'elanger, and
  Boehm}}]{AlbornozVasquez:2011js}
\bibinfo{author}{\bibfnamefont{D.}~\bibnamefont{Albornoz~V\'asquez}},
  \bibinfo{author}{\bibfnamefont{G.}~\bibnamefont{B\'elanger}},
  \bibnamefont{and} \bibinfo{author}{\bibfnamefont{C.}~\bibnamefont{Boehm}},
  \bibinfo{journal}{Phys.Rev.} \textbf{\bibinfo{volume}{D84}},
  \bibinfo{pages}{095008} (\bibinfo{year}{2011}), \eprint{1107.1614}.

\end{thebibliography}

\end{document}